\documentclass[aps,pra,amsmath,reprint,floatfix,superscriptaddress]{revtex4-2}
\usepackage{microtype} 
\usepackage{amssymb}
\usepackage{graphicx}
\usepackage[usenames,dvipsnames]{xcolor} 
\usepackage{bm}
\usepackage{times} 
\allowdisplaybreaks 

\renewcommand{\vec}[1]{{\bm{#1}}} 
\renewcommand{\Re}{\mathrm{Re}\!\;}
\renewcommand{\Im}{\mathrm{Im}\!\;}
\newcommand{\md}{\mathrm{d}}
\newcommand{\mK}{{\mathrm{K}}}
\newcommand{\mHK}{{\mathrm{HK}}}
\newcommand{\mDN}{{\mathrm{DN}}}
\newcommand{\msp}{{\mathrm{sp}}}
\newcommand{\morb}{{\mathrm{orb}}}
\newcommand{\Ac}{\mathcal{A}}
\newcommand{\mi}{\mathrm{i}}
\newcommand*{\ndots}{\kern-0.15em.\kern-0.05em.\kern-0.05em.} 

\RequirePackage[
  hyperindex,colorlinks,bookmarksnumbered,
  plainpages=true,pdfstartview=FitH]{hyperref}
\hypersetup{urlcolor={blue!50!black}, linkcolor={red!50!black},citecolor={blue!50!black}} 
\usepackage{hyperref}
\usepackage[all]{hypcap}

\begin{document} 

\title{Low-energy perspective on two-orbital Hund metals and the case of LaNiO\texorpdfstring{$_2$}{2}}
\author{Fabian B.~Kugler}
\affiliation{Center for Computational Quantum Physics, Flatiron Institute, 162 5th Avenue, New York, NY 10010, USA}
\affiliation{Department of Physics and Astronomy and Center for Materials Theory, Rutgers University, Piscataway, NJ 08854, USA}
\author{Chang-Jong Kang}
\affiliation{Department of Physics, Chungnam National University, Daejeon 34134, South Korea}
\affiliation{Department of Physics and Astronomy and Center for Materials Theory, Rutgers University, Piscataway, NJ 08854, USA}
\author{Gabriel Kotliar}
\affiliation{Department of Physics and Astronomy and Center for Materials Theory, Rutgers University, Piscataway, NJ 08854, USA}
\affiliation{Condensed Matter Physics and Materials Science Department,\looseness=-1\,  
Brookhaven National Laboratory, Upton, NY 11973, USA}
\date{\today}

\begin{abstract}
The Hund-metal route to strong correlations continues to attract large interest in the condensed-matter community. The question arose to what extent it applies to the infinite-layer nickelates and, as a related question, to two-orbital systems in general. Here, we provide a low-energy perspective on this topic through a dynamical mean-field study using the numerical renormalization group (NRG) as a real-frequency impurity solver. We find that the RG flow from high to low energy is a uniquely adequate tool to reveal two-stage Kondo screening (2SKS), a fascinating mechanism for Hund physics. Further, we show that 2SKS takes place in a quarter-filled two-orbital system, but can be easily suppressed by a sufficiently large crystal-field splitting. We apply these insights to LaNiO$_2$ using a recently proposed two-orbital model and show that it is indeed the crystal-field splitting that suppresses multiorbital phenomena in this scenario. Our general findings open the way for further explorations of 2SKS, and we propose a way of potentially inducing low-energy Hund physics in LaNiO$_2$ by counteracting the crystal field.
\end{abstract}

\maketitle

\section{introduction}
Recently, the physics of Hund metals \cite{Yin2011}, 
where not only Hubbard $U$ but also Hund's coupling $J$ plays a key role for the electronic correlations,	
is discussed in a broader context than originally anticipated.
Archetypal Hund metals (see Ref.~\cite{Georges2013,Georges2024} for reviews) are
iron pnictides and chalcogenides as well as ruthenates,
whose occupancy of the $d$ shell is away from half-filling or single occupancy.
However, it was also proposed 
that Hund physics is relevant for the half-filled NiS$_{2-x}$Se$_x$ \cite{Jang2021,Park2024}
or for the infinite-layer nickelates (which bear similarities to the cuprates).

Indeed, the finding of unconventional superconductivity 
in the nickelates \cite{Li2019} 
(see Ref.~\cite{Review2024,Review2022,Review2022-1,Review2022-2} for reviews)
has revived the theoretical interest in these compounds.
There are by now many works 
using the combination of 
density functional theory and dynamical mean-field theory (DFT+DMFT)
\cite{Georges1996,Kotliar2006}
on 112 nickelates (such as NdNiO$_2$ and LaNiO$_2$) 
\cite{Karp2020,Lechermann2020,Gu2020,Ryee2020,Si2020,Kitatani2020,%
Lechermann2020b,Leonov2020,Wang2020,Kang2021,Kang2023}.
Some of them emphasize more the single-orbital nature of correlations with an additional self-doping band  \cite{Karp2020,Kitatani2020}, 
others evoke a  multiorbital Hund-metal picture \cite{Wang2020,Kang2023}.
Recently \cite{Kang2021}, it was pointed out that, while the $d_{x^2-y^2}$ orbital dominates the low-energy properties, there can be important orbital fluctuations at intermediate energies.

Understanding the influence of Hund physics in 112 nickelates,
especially within their Ni $e_g$ orbitals,
requires answering a simpler and more general question: To what extent can Hund physics be present in two-orbital systems,
where nonzero integer occupancy is bound to either half-filling or single occupancy (of an electron or a hole)?
Addressing this question, Ref.~\cite{Ryee2021} studied a quarter-filled two-orbital model,
finding that one can indeed distinguish a \textit{weak Hund metal}.
This study employed a quantum Monte Carlo imaginary-frequency DMFT impurity solver \cite{Gull2011,Choi2019}.
Here, we confirm this finding and provide an additional perspective 
by using the numerical renormalization group (NRG) \cite{Bulla2008}
as a real-frequency impurity solver,
which offers exponentially fine low-energy resolution and 
a physically transparent RG flow.

The goal of this paper is twofold.
First, we take a minimal-model point of view
and analyze an orbital-symmetric two-orbital system with occupancy 1,
as compared to three-orbital systems with occupancy 1 or 2.
Through properties not studied in Ref.~\cite{Ryee2021} (real-frequency spectra and RG flows), we substantiate their finding of a weak Hund metal.
Compared to related DMFT+NRG works \cite{Stadler2015,Stadler2019,Kugler2019,Walter2020}, our juxtaposition shows in detail 
how the different Hund-metal characteristics play out in the different scenarios 
and reveals the two-orbital weak Hund metal not previously appreciated with this method.
We also scrutinize the influence of a crystal-field splitting between the two orbitals and show how it suppresses Hund-metal phenomena.
Second, we turn to a material-realistic setting
and analyze a two-orbital description of LaNiO$_2$ 
\cite{Sakakibara2020,Hepting2020,Adhikary2020,Wan2021,Kang2021}
at quarter-filling.
We show that---in this specific scenario---multiorbital Hund physics is suppressed by a large crystal-field splitting of $\Delta \!\approx\! 0.8 \,\mathrm{eV}$.
This is evidenced by comparing low-energy properties of the $d_{x^2-y^2}$ orbital obtained from a two- and a one-orbital calculation
and by observing Hund physics at $\Delta \!=\! 0$.

The rest of the paper is organized as follows.
The models and the method are briefly introduced in Sec.~\ref{sec:models_method}.
The general model analysis is contained in Secs.~\ref{sec:general_considerations}--\ref{sec:2orb_cfs} and the LaNiO$_2$ study in Sec.~\ref{sec:LaNiO2}.
We conclude in Sec.~\ref{sec:conclusion}.
Appendix~\ref{app:more_results} contains additional numerical results and App.~\ref{app:Hamiltonians}
general statements about interaction Hamiltonians.

\section{Models and method}
\label{sec:models_method}
We consider multiorbital Hubbard models of the type
\begin{align}
H
=
\sum_{ijmm'\sigma} 
d^\dag_{im\sigma} h^{mm'}_{ij} d_{jm'\sigma} 
+
\sum_i
H_\mathrm{int}[d_{im\sigma}]
,
\label{eq:Hamiltonian}
\end{align}
where 
$d^\dag_{im\sigma}$
creates an electron at site $i$, in orbital $m \!\in\! \{ 1, \ndots, \! M \}$, and with spin $\sigma \!\in\! \{ \uparrow, \downarrow \}$.
For the first part of our results, the hopping matrix is taken to be diagonal in orbital space
and such that it corresponds to a semicircular density of states of half bandwidth $D \!=\! 1$ 
(Bethe lattice). 
We consider both the orbital-symmetric case with on-site energies $\epsilon_m \!=\! \epsilon_{m'}$ for two and three orbitals as well as two-orbital systems with a crystal-field splitting $\Delta \!=\! \epsilon_2 \!-\! \epsilon_1$.
The three-orbital case with occupancy 2 serves as our reference system for ``canonical'' Hund physics
\cite{Georges2013,Stadler2015,Ryee2023}.
In the second part of our results, the hopping matrix is taken from a two-orbital model fitted to the band structure of LaNiO$_2$.
It thus has off-diagonal elements in orbital space, except for the local part $i \!=\! j$.
We remark that three-orbital models of LaNiO$_2$ were suggested too \cite{Nomura2019,Liu2020,Karp2020},
but these are not considered in the present work.

The local interaction is chosen as ($n_{m\sigma} = d^\dag_{m\sigma} d_{m\sigma}$)
\begin{align}
H_{\mathrm{int}}
& =
U \sum_m n_{m\uparrow} n_{m\downarrow}
+
\!\sum_{m<m',\sigma\sigma'}\!
( U' - \delta_{\sigma\sigma'} J )
n_{m\sigma} n_{m'\sigma'}
\nonumber \\
& \ -
J \sum_{m\neq m'} d^\dag_{m\uparrow} d_{m\downarrow} d^\dag_{m'\downarrow} d_{m'\uparrow}
.
\end{align}
We show in App.~\ref{app:Hamiltonians} that,
at $U' \!=\! U \!-\! J$, this realizes the Dworin--Narath (DN) Hamiltonian
$(U - \tfrac{3}{2}J) \tfrac{1}{2}N(N \!-\! 1) - J \vec{S}^2 + \tfrac{3}{4} J N$.
At $U' \!=\! U \!-\! 2J$, it corresponds to the Hubbard--Kanamori (HK) Hamiltonian
$(U-3J) \tfrac{1}{2}N(N \!-\! 1) - 2J \vec{S}^2 - \tfrac{1}{2} J \vec{L}^2 + \tfrac{2+M}{2} J N$
($\vec{L} \!\equiv\! L_3$ for $M \!=\! 2$),
except that pair-hopping is neglected.
The latter restriction does not qualitatively change the low-energy Hund physics but simplifies the numerics considerably \cite{Kugler2020}.
When ambiguous, we attach a subscript to the Hund's coupling, $J_\mDN$ or $J_\mHK$, to specify which Hamiltonian is being used.

We follow Ref.~\cite{Ryee2021} in defining an average (opposite-spin) Hubbard interaction $\bar{U} \!=\! 
\frac{1}{M^2} \sum_{mm'} U_{mm'}$ (called $U_{\mathrm{av}}$ in Ref.~\cite{Ryee2021}), where
$U_{mm} \!=\! U$ and $U_{m\neq m'} \!=\! U'$. This yields
\begin{align}
\bar{U}
= 
U' \!+\! \frac{U-U'}{M}
=
\begin{cases}
U \!-\! \frac{M-1}{M}J_\mDN, \!\! & U' = U \!-\! J_\mDN
\\
U \!-\! 2\frac{M-1}{M}J_\mHK, \!\! & U' = U \!-\! 2J_\mHK
\end{cases}
.
\end{align}
The cases relevant for us are listed in Table~\ref{tab:Uav_DeltaAt},
along with the corresponding values of the atomic gap $\Delta_{\mathrm{at}} \!=\! U' \!-\! J$ \cite{Georges2013}.
\begin{table}
\centering
\begin{tabular}{c|c|c|c|c}
	& $M\!=\!2 $ DN & $M\!=\!2 $ HK & $M\!=\!3 $ DN & $M\!=\!3 $ HK
\\ \hline
$\bar{U}$ & $U\!-\!\tfrac{1}{2}J$ & $U\!-\! J$ & $U\!-\!\tfrac{2}{3}J$  & $U\!-\!\tfrac{4}{3}J$
\\ \hline
$\Delta_{\mathrm{at}}\!=\! U'\!-\! J$ & $U\!-\! 2J$ & $U\!-\! 3J$ & $U\!-\! 2J$  & $U\!-\! 3J$
\end{tabular}
\caption{Average (opposite-spin) Hubbard interaction $\bar{U}$ and atomic gap $\Delta_{\mathrm{at}}$ for $M \!=\! 2$ or $3$ orbitals using the DN or HK Hamiltonian.}
\label{tab:Uav_DeltaAt}
\end{table}

\begin{table}
\centering
\begin{tabular}{c|c|c|c|c}
$M$ & $H_{\mathrm{int}}$ & Symmetry & $N_{\mathrm{kp}}$ & $N_{\mathrm{kp}}^*$
\\ \hline
3 & DN
& 
$\textrm{U}(1)_{\textrm{ch}} \otimes \textrm{SU}(2)_{\textrm{sp}} \otimes \textrm{SU}(3)_{\textrm{orb}}$
& 
$20\textsf{k}$ & $800\textsf{k}$
\\ \hline
2 & DN
& 
$\textrm{U}(1)_{\textrm{ch}} \otimes \textrm{SU}(2)_{\textrm{sp}} \otimes \textrm{SU}(2)_{\textrm{orb}}$
& 
$15\textsf{k}$ & $130\textsf{k}$
\\ \hline
3 & HK
& 
$\bigotimes_{m=1}^3 \textrm{U}(1)_{\textrm{ch},m} \otimes \textrm{SU}(2)_{\textrm{sp}}$
& 
$100\textsf{k}$ & $350\textsf{k}$
\\ \hline
2 & HK
& 
$\bigotimes_{m=1}^2 \textrm{U}(1)_{\textrm{ch},m} \otimes \textrm{SU}(2)_{\textrm{sp}}$
& 
$30\textsf{k}$ & $90\textsf{k}$
\end{tabular}
\caption{Symmetries used and multiples $N_{\mathrm{kp}}$ (or states $N_{\mathrm{kp}}^*$) maximally kept during the NRG iterative diagonalization ($1\textsf{k}$ denotes $10^3$).}
\label{tab:NRG_setting}
\end{table}

We treat our models within the DMFT approximation \cite{Georges1996,Kotliar2006}, 
where the lattice Hamiltonian is mapped onto a self-consistently determined impurity model. 
The latter is then solved with NRG \cite{Bulla2008} at zero temperature ($T \!=\! 10^{-6}$ or $10^{-8}$ in practice) assuming paramagnetism. 
We employ the full density-matrix NRG \cite{Peters2006,Weichselbaum2007}
in a state-of-the-art implementation based on the QSpace tensor library \cite{Weichselbaum2012a,*Weichselbaum2012b,*Weichselbaum2020}.
We use an adaptive broadening scheme \cite{Lee2016,Lee2017}
and a symmetric improved estimator for the self-energy \cite{Kugler2022}.
In the cases without $\textrm{SU}(M)$ orbital symmetry, we interleave the Wilson chains for increased efficiency \cite{Mitchell2014,Stadler2016}.
For the overview of quasiparticle weights, we used $n_z \!=\! 2$ shifted discretization grids \cite{Zitko2009}; for all other calculations, $n_z \!=\! 4$.
We purposefully pick the same NRG discretization parameter $\Lambda \!=\! 4$ for \textit{all} calculations, in order to have the most straightforward comparison between different flow diagrams.
All available \textbf{ch}arge, \textbf{sp}in, and \textbf{orb}ital symmetries are exploited,
as summarized in Table~\ref{tab:NRG_setting}.
There, we also give the maximal number of multiplets $N_{\mathrm{kp}}$ (and corresponding number states $N_{\mathrm{kp}}^*$) kept during the iterative diagonalization.

The quantities of interest are the
(retarded) local self-energies $\Sigma_m(\omega)$ 
and associated quasiparticle weights $Z_m \!=\! [1 \!- \!\partial_\omega \Re \Sigma_m(\omega)|_{\omega=0}]^{-1}$,
the local spectral functions $\Ac_m(\omega)$
and associated occupancy per orbital $n_m \!=\! \int \md \omega f_\omega \Ac_{\mathrm{loc},m}(\omega)$ (where $f_\omega \!=\! [1 + \exp(\omega/T)]^{-1}$),
and the (retarded) local susceptibilities $\chi \!=\! \chi' \!-\! \mi\pi\chi''$
defined via the spin and orbital operators 
(cf.\ App.~\ref{app:Hamiltonians}).
We further analyze NRG flow diagrams \cite{Bulla2008}
(see Ref.~\cite{Kugler2020} for how $E_\ell$ is mapped to $\omega$).
Note that, for the models solved with interleaved NRG,
we perform an additional diagonalization without interleaving 
in order to have flow diagrams comparable to those from the other models.
Finally, in our LaNiO$_2$ study, we also consider the momentum-resolved 
spectral function
$\Ac(\vec{k},\omega) \!=\! -\frac{1}{\pi} \Im \mathrm{Tr} [\omega \!+\! \mu \!-\! \vec{h}(\vec{k}) \!-\! \vec{\Sigma}(\omega)]^{-1}$, following from the matrix-valued hopping and self-energy.

\section{Results}
\subsection{General considerations}
\label{sec:general_considerations}

\begin{figure}[t]
\includegraphics[scale=1]{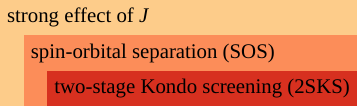}
\caption{%
Venn diagram showing various effects of Hund's coupling $J$.
The arguably most intricate effect is two-stage Kondo screening (2SKS):
upon decreasing energy, first, orbital fluctuations are dynamically (Kondo-) screened, then spin fluctuations.
This phenomenon was dubbed spin-orbital separation (SOS) \cite{Stadler2015}, with $T^\mK_\msp \!\ll\! T^\mK_\morb \!\ll\! U,J$.
However, SOS can also occur more trivially.
In a half filled system at finite $J$, e.g., orbital fluctuations are suppressed at energies below $J$
(or, in systems with a crystal field $\Delta$, at energies below $\Delta$ \cite{Kugler2019}).
If, in abuse of notation, one still calls this energy scale $T^\mK_\morb$, one then has
$T^\mK_\msp \!\ll\! T^\mK_\morb \!\sim\! J,\Delta$.
Finally, $J$ not only affects low-energy Kondo physics but also (high-energy) atomic physics,
like orbital occupancies.
Thereby, it, e.g., facilitates an orbital-selective Mott phase in systems with broken orbital symmetry \cite{Medici2011a} (see also Ref.~\cite{Kugler2022a}).}
\label{fig:Jclass}
\end{figure}

\begin{figure*}[t]
\includegraphics[scale=1]{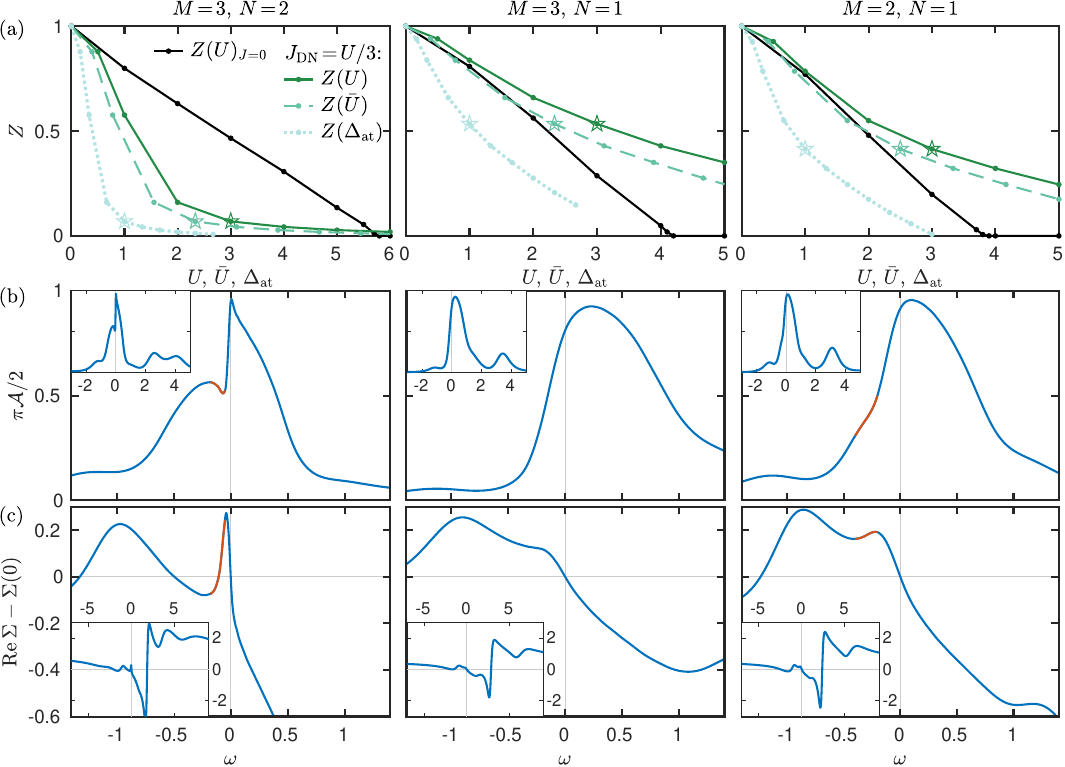}
\caption{%
(a) Quasiparticle weight $Z$ as a function of $U$, $\bar{U}$, or $\Delta_{\mathrm{at}}$. The Janus effect ($Z|_{J\neq 0}\!<\! Z|_{J=0}$ for small $U$, $Z|_{J\neq 0}\!>\! Z|_{J=0}$ for large $U$) is strikingly clear for $(M,N) \!=\! (3,2)$ and somewhat discernible for $(M,N) \!=\! (2,1)$ when going from $Z(U)$ to $Z(\bar{U})$.
(b) Spectral functions. The red line indicates the orbital-resonance shoulder.
(c) Real part of self-energies. The red line indicates the corresponding inverted slope.}
\label{fig:Z_A_SE}
\end{figure*}

Hund's coupling $J$ can affect interacting systems in various ways.
For instance, $J$ changes the energies of atomic configurations and thereby influences their relative probabilities.
In both the DN and the HK Hamiltonians, $J$ acts more strongly (with a larger prefactor) on the spin than on the orbital operators squared.
Thereby, it naturally induces an asymmetry between the spin and orbital sectors.
In the study of Kondo scales $T^\mK$, indicating below which energies spin and orbital fluctuations are dynamically screened, 
this asymmetry with $T^\mK_\msp \!\ll\! T^\mK_\morb$ was dubbed spin-orbital separation (SOS) \cite{Stadler2015,Stadler2019}.
Besides the Kondo screening mechanism, fluctuations can also be quenched more trivially, say, by atomic energy considerations.
In a half-filled system at finite $J$, e.g., the atomic ground state is an orbital singlet, and orbital fluctuations are suppressed at energies below the bare parameter $J$.
Similarly, in systems with large crystal-field splitting $\Delta$, orbital fluctuations are suppressed at energies below the bare parameter $\Delta$.
The various aspects of $J$ mentioned above are illustrated in Fig.~\ref{fig:Jclass}.
In the following, we focus on one particular aspect: the highly nontrivial form of SOS induced by two-stage Kondo screening (2SKS).

How does the effect of $J$, and 2SKS in particular, depend on the number of orbitals $M$ and the occupancy $N$?
From an atomic perspective, $J$ splits the manifold of energy eigenstates with two or more electrons, giving preference to the high-spin state. 
Accordingly, this directly affects the atomic ground state for $N \!\geq\! 2$, while it only affects excited states for $N \!=\! 1$.
Going beyond the atomic picture, 
early works \cite{Yin2012,Aron2015} utilized a multiorbital Schrieffer--Wolff transformation.
The spin Kondo coupling in a one-orbital system is
$J^\mK_\msp \!=\! 2V^2 ( \frac{1}{\Delta E_-} \!+\! \frac{1}{\Delta E_+} )$
where $\Delta E_{\pm} \!>\! 0$ is the energy difference to the states with charge $\pm 1$ w.r.t.\ the ground state.
In the one-orbital example, these are $\Delta E_+ \!=\! \epsilon \!+\! U$ and $\Delta E_+ \!=\! -\epsilon$,
whence the well-known result $J^\mK_\msp \!=\! 2V^2 \frac{U}{(\epsilon+U)(-\epsilon)}$ 
and $J^\mK_\msp \!=\! 8V^2/U$ at the particle-hole symmetric point $\epsilon \!=\! -U/2$ follows.
In multiorbital systems (without loss of generality $M \!>\! N$), it was found \cite{Yin2012,Aron2015} that
$J^\mK_\msp \!=\! \frac{2}{M} V^2 ( \frac{1}{\Delta E_-} - \frac{M-N}{N+1} \frac{1}{\Delta E_+} )$.
By the negative sign, $J^\mK_\msp$ is reduced, thereby suppressing the spin Kondo temperature.
(Note that an integer occupancy is achieved by balancing charge fluctuations to the adjacent sectors,
so often times $\Delta E_- \!\approx\! \Delta E_+$.)
Importantly, however, this statement applies to all values of $M$ and $N \!<\! M$. 
Hence, in contrast to the atomic perspective, it does not give preference to Hund physics in, say, systems with $(M,N) \!=\! (3,2)$ over those with $(M,N) \!=\! (2,1)$.

\begin{figure*}[t]
\includegraphics[scale=1]{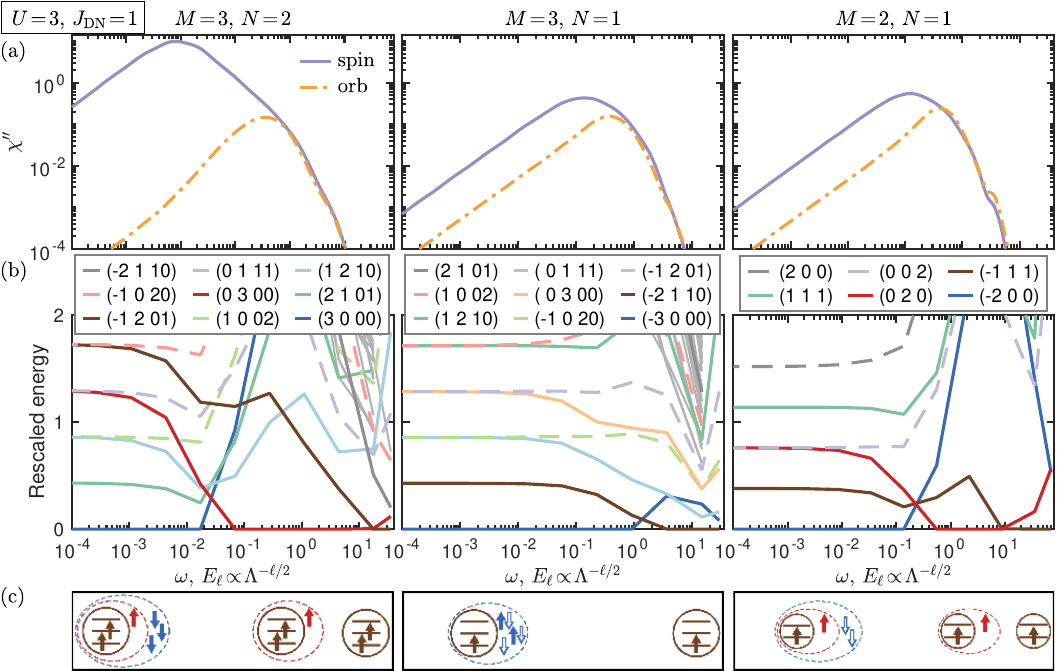}
\caption{%
(a) Susceptibilities showing SOS: $\chi''_{\mathrm{sp}}$ is peaked at higher frequencies than $\chi''_\morb$.
(b) NRG flow diagrams. The legend gives the charge (with respect to half-filling), spin, and orbital quantum numbers.
(c) Pictorial representation of the ground states during the flow.
The systems at charge one away from half-filling can realize an intermediate orbital-singlet maximal-spin state (red) at the lowest energy and thus experience 2SKS.}
\label{fig:chi_flow}
\end{figure*}

\subsection{Single-particle properties in orbital-symmetric models}
Let us now turn to numerical results.
For the orbital-symmetric models, we employ the DN Hamiltonian,
but the same behavior is observed for the HK Hamiltonian, see App.~\ref{app:more_results}.
Figure~\ref{fig:Z_A_SE}(a) shows how $Z$ depends on the interaction strength in various settings
and three filling cases: $(M,N) \!\in\! \{ (3,2), (3,1), (2,1) \}$.
First, increasing $U$ at $J \!=\! 0$ (black lines), one finds a Mott insulator for sufficiently large $U$ in all three cases.
Finite $J$ changes the behavior drastically, see the (dark) green line for $Z(U)$ at $J \!=\! U/3$.

The left panel $(M,N) \!=\! (3,2)$ displays the familiar Janus effect \cite{Medici2011,Georges2013}:
compared to the black curve, 
$Z$ at finite $J$ drops down for small $U$ and then flattens again for larger $U$, so that $Z(U)|_{J\neq 0}$ exceeds $Z(U)|_{J=0}$ for $U \!\gtrsim\! 6$.
The latter property is a rather trivial atomic effect, as $J$ reduces the atomic gap $\Delta_{\mathrm{at}} \!=\! U' \!-\! J$ away from half-filling
\cite{Medici2011,Georges2013}.
Indeed, this effect disappears if replacing $U$ by $\Delta_{\mathrm{at}}$ as the independent variable, and one has
$Z(\Delta_{\mathrm{at}})|_{J\neq 0} \!<\! Z(U \!=\! \bar{U} \!=\! \Delta_{\mathrm{at}})|_{J = 0}$ throughout.
By contrast, the sharp drop of $Z$ at relatively small $U$ is a manifestation of Hund physics,
i.e., strong correlations induced by Hund-$J$ at moderate Hubbard-$U$.
As explained in Ref.~\cite{Medici2011}, this is attributed to lifting the degeneracy of the atomic ground state (from 15-fold to ninefold degenerate \cite{Georges2013}).

In the middle panel, $(M,N) \!=\! (3,1)$, there is no such Janus effect. 
Instead, finite $J$ increases $Z(U)$ for all $U$ (an effect that is again eliminated by switching from $Z(U)$ to $Z(\Delta_{\mathrm{at}})$).
The right panel, $(M,N) \!=\! (2,1)$, also has $Z(U)|_{J\neq 0} \!>\! Z(U)|_{J=0}$.
Perhaps, one might say that the initial decrease and subsequent flattening of $Z(U)_{J\neq 0}$ is slightly more pronounced 
in the $(M,N) \!=\! (2,1)$ than in the $(M,N) \!=\! (3,1)$ case.

In Ref.~\cite{Ryee2021}, it was argued that $\bar{U}$ instead of $U$ provides a better reference on the horizontal axis.
Going from $Z(U)$ to $Z(\bar{U})$ shifts the $Z|_{J\neq 0}$ curve to the left by a relatively small amount (compared to going from  $Z(U)$ to $Z(\Delta_{\mathrm{at}})$)
proportional to $J$ (and thus $U$),
see the dashed line in Fig.~\ref{fig:Z_A_SE}(a) and Table~\ref{tab:Uav_DeltaAt}.
The qualitative behavior in the left and middle panels of Fig.~\ref{fig:Z_A_SE}(a) remains basically unchanged.
For the right panel, one might now argue that $Z(\bar{U})|_{J\neq 0}$ initially drops to an appreciable extent below $Z(U \!=\! \bar{U})|_{J=0}$
and then crosses through $Z(U \!=\! \bar{U})|_{J=0}$ due to a moderate flattening.
In this way, one may say that $J$ slightly increases the correlation strength at $\bar{U} \!\sim\! 1$,
and, thus, the system for $(M,N) \!=\! (2,1)$ constitutes a \textit{weak} Hund metal \cite{Ryee2021}.
However, since the splitting of atomic energy manifolds by $J$ affects only the excited states for $(M,N) \!=\! (2,1)$ (and not the ground state, as for $(M,N) \!=\! (3,2)$),
the drop in $Z$ is much smaller for $(M,N) \!=\! (2,1)$ than for $(M,N) \!=\! (3,2)$.

\begin{figure*}[t]
\includegraphics[scale=1]{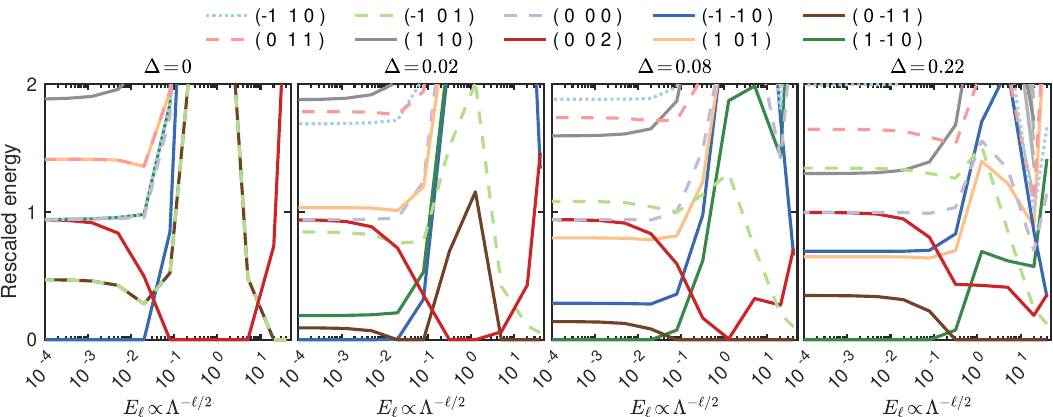}
\caption{%
NRG flow diagrams at variable crystal-field splitting $\Delta$.
The intermediate spin-triplet ground state 
(present in regions where the red line reaches zero)
is only found for $|\Delta| \!\lesssim\! T^\mK_\morb(\Delta \!=\! 0)/2$.}
\label{fig:Deltaflow}
\end{figure*}

Next, Figs.~\ref{fig:Z_A_SE}(b) and \ref{fig:Z_A_SE}(c) show the spectral function and the real part of the self-energy in each case at $U \!=\! 3$ and $J \!=\! 1$ (the point indicated by a star in Fig.~\ref{fig:Z_A_SE}(a)).
The insets in Fig.~\ref{fig:Z_A_SE}(b) display $\Ac$ on a wide-energy window, while the main plots focus on the quasiparticle peak.
An intriguing feature observed in Hund metals is a orbital-resonance shoulder 
\cite{Stadler2015,Stadler2019,Kugler2019}
in the left part (for $N \!<\! M$) of the quasiparticle peak.
Indeed, such a shoulder (marked in red) is very pronounced for $(M,N) \!=\! (3,2)$ and somewhat visible for $(M,N) \!=\! (2,1)$.
In $\Re \Sigma$, this corresponds to an inverted slope or retracted renormalization (marked in red)
\cite{Iwasawa2005,Mravlje2011,Stricker2014,Wadati2014,Kugler2019,Kugler2020,Karp2020b}. 
The inverted slope is also strikingly clear for $(M,N) \!=\! (3,2)$ and only somewhat discernible for $(M,N) \!=\! (2,1)$.
Nevertheless, the fact that both features occur for $(M,N) \!=\! (3,2)$ and (albeit less pronounced) for $(M,N) \!=\! (2,1)$
corroborates the notion of a \textit{weak} Hund metal in the latter case.
Note that they are also found in rather different settings.
For instance, as elaborated in Ref.~\cite{Kugler2019}, they persist when introducing a crystal-field splitting $\Delta$. Upon increasing $\Delta$ and approaching an orbital-selective Mott phase, 
they eventually change in nature from an orbital resonance (corresponding to 2SKS) 
to a separate doublon-holon peak (atomiclike).
The same behavior occurs for the two-orbital setting with crystal-field splitting analyzed below.

\subsection{NRG flow in orbital-symmetric models}
We now inspect SOS and 2SKS via the susceptibilities and the NRG flow diagrams in each case.
From $\chi''_{\mathrm{sp}}$ and $\chi''_\morb$ in Fig.~\ref{fig:chi_flow}(a), one readily observes SOS \cite{Stadler2016,Stadler2019}:
the peak of the orbital susceptibility (indicating that orbital fluctuations get screened) occurs at a higher energy than the peak of the spin susceptibility (indicating that spin fluctuations get screened).
This phenomenon is very pronounced for $(M,N) \!=\! (3,2)$ and weakly present in the other two cases.
Whether 2SKS causes SOS is revealed by the NRG flow.

In Figs.~\ref{fig:chi_flow}(b) and \ref{fig:chi_flow}(c), we present the NRG flow diagrams (to be read from right to left) and 
a pictorial representation of the ground state in different parts of the RG flow,
following Refs.~\cite{Stadler2019,Walter2020}.
The legend in Fig.~\ref{fig:chi_flow}(b) gives the U(1) charge (relative to half-filling), SU(2) spin (twice $S$) and SU($M$) orbital quantum numbers.
At the beginning of the flow (i.e., at high energy, to the right in each panel of Fig.~\ref{fig:chi_flow}(b)), 
the system is fully described by an atomic picture.
There, the ground state (brown line) can be represented by $N$ (aligned) spins symmetrically placed in $M$ orbitals.
Going to lower energies (leftward in the plots), the effect of the environment (the self-consistently determined bath) comes into play.

The left panel, $(M,N) \!=\! (3,2)$, is our reference system for Hund physics and 2SKS.
One clearly observes an extended region of the flow where the lowest-energy state (red line) involves the impurity with an additional bath electron (charge increased by one) that screens the orbital moment (quantum number $00$) and further increases the spin (quantum number $3$). 
This is part 1 of 2SKS.
At the end of the flow (at low energy, to the left of the plot), three more bath electrons subsequently screen the magnetic moment (blue line; charge increased by 3, zero spin and orbital quantum numbers) \cite{Stadler2019,Walter2020}.
This is part 2 of 2SKS, yielding the Fermi liquid.

For $(M,N) \!=\! (3,1)$, such a 2SKS is not observed. As the system is further away from half-filling, an intermediate orbital-singlet and spin-tripled state would require the impurity to temporarily bind two bath electrons. Instead, the system directly flows from the atomic state (brown line, one electron in three orbitals) to a fully screened ground state (blue line).

Now, for $(M,N) \!=\! (2,1)$, again one away from half-filling, we do observe 2SKS: 
first, the orbital moment of one electron in two orbitals (brown line) is screened by one bath electron. The resulting spin triplet (red line) is then screened by two bath holes, leading to the fully screened Fermi-liquid ground state (blue line). 
Whether the final screening is achieved by electrons or holes depends on the detailed shape of the particle-hole asymmetric hybridization function at low energies.
As before, Hund-metal signatures (here 2SKS) are observed for three and two orbitals at charge one away from half-filling, but much weaker so for $M \!=\! 2$ than for $M \!=\! 3$.

We end this section with a note of caution.
The RG flow in NRG is discrete, owing to the logarithmic discretization with $\Lambda \!>\! 1$.
Even--odd oscillations and large $\Lambda$ values needed in multiorbital systems
make the RG flow rather coarse.
Depending on how the NRG parameters ($\Lambda$, $z$ shifts, etc.) are chosen and how close the putative RG fixed point is to the given RG trajectory, one may or may not explicitly see it.
Here, for comparing different filling cases, we chose $\Lambda \!=\! 4$ throughout
and were indeed able to see the intermediate maximal-spin state.

\subsection{NRG flow in two-orbital model with crystal-field splitting}
\label{sec:2orb_cfs}
Restricting ourselves from now on to $(M,N) \!=\! (2,1)$,
we next incorporate the effect of crystal-field splitting.
We stick to the previous setting with a semicircular density of states but
allow for a finite $\Delta \!=\! \epsilon_2 \!-\! \epsilon_1$.
Since we thereby break orbital symmetry, we also exclusively use the HK Hamiltonian henceforth.
As interaction values, we choose $U \!=\! 3.1$ and $J \!=\! 0.7$ in units of $D \!=\! 1$.
These are the same values chosen later for the two-orbital model of LaNiO$_2$ (then in units of eV).

In Ref.~\cite{Kugler2019}, it was explained how the orbital-resonance shoulder is replaced by a separate peak if $|\Delta| \!\gtrsim\! T^\mK_\morb(\Delta \!=\! 0)/2$.
Here, we use the same line of reasoning but focus on the NRG flow.
Figure~\ref{fig:Deltaflow} shows the flow diagrams for four values of $\Delta$;
the legend gives the two U(1) charges per orbital (with respect to a half-filled orbital) and the SU(2) spin quantum number (twice $S$).
At the orbital-symmetric point, $\Delta \!=\! 0$, the flow is similar to the leftmost panel in Fig.~\ref{fig:chi_flow}(b):
it starts from the atomic ground state (degenerate brown and light green lines) and goes through a regime where the lowest-energy state has an additional bath electron screening the orbital momentum and forming a spin triplet (red line). The flow ends at the fully screened Fermi liquid where the ground state has two additional holes (one per orbital) screening the magnetic moment (blue line).

Small $\Delta \!=\! 0.02$ breaks the orbital symmetry, and the lines split accordingly, but the same three lowest-energy states in the three regions persist.
At larger $\Delta \!=\! 0.08$, the lines shift further compared to the previous two cases. 
Most importantly, the intermediate spin-triplet state (red line) barely touches zero energy 
and therefore has a much weaker effect during the RG flow. 
The screening of the magnetic moment,
yielding the Fermi-liquid ground state (dark green line)
now occurs via an electron in the lower-in-energy orbital (its charge is increased by one).
Finally, at very large $\Delta \!=\! 0.22$, there is no notion of 2SKS anymore. 
States involving the orbital higher in energy have shifted away from zero energy, 
and the standard, one-stage Kondo screening occurs exclusively in the orbital
lower in energy (brown line to dark green line).

If one merely looks at the susceptibilities,
it might appear that the SOS window has actually increased from $\Delta \!=\! 0$ to $\Delta \!=\! 0.22$.
The reason is that, for large $\Delta$, $\chi''_\morb$ roughly peaks at $\omega \!\sim\! \Delta$ \cite{Kugler2019}.
This, however, does not necessarily mean one is ``deeper'' in the Hund-metal regime.
Instead, the mechanism for SOS has changed,
from 2SKS to atomic energy considerations.
The same is true for the shoulder in the spectral function and the inverted slope in the self-energy. With increasing $\Delta$, both features become more pronounced in the orbital pushed to half-filling (and less pronounced in the other orbital). 
For large $\Delta$, however, they no longer stem from 2SKS but from doublon-holon subpeaks \cite{Kugler2019}.

\subsection{Two-orbital model of LaNiO\texorpdfstring{$_2$}{2}}
\label{sec:LaNiO2}

\begin{figure}[t]
\includegraphics[scale=1]{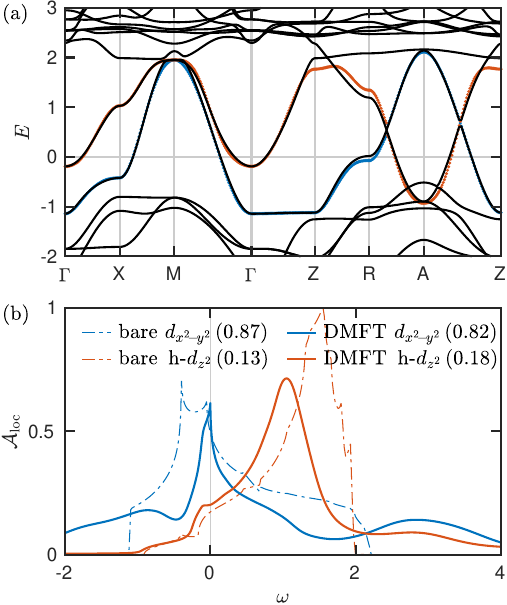}
\caption{%
(a) Band structure of LaNiO$_2$ from DFT (black)
and our two-orbital model ($d_{x^2-y^2}$ in blue and $\textrm{h-}d_{z^2}$ in red).
(b) Local spectral functions from (a) (bare) and DMFT ($U \!=\! 3.1$, $J \!=\! 0.7$).
The occupancy per orbital is given in brackets.}
\label{fig:LaNiO2_spectrum}
\end{figure}

\begin{figure*}[t]
\includegraphics[scale=1]{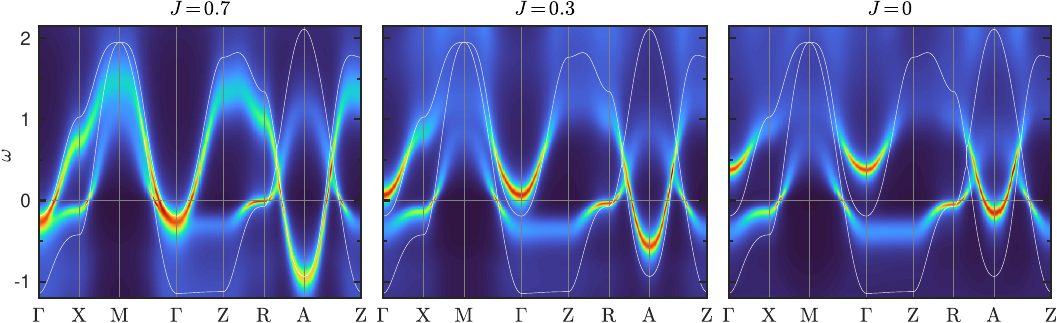}
\caption{%
Momentum-resolved spectral function $\mathcal{A}({\vec{k},\omega})$ of LaNiO$_2$ at variable $J$ and fixed $U \!=\! 3.1$ (with a broadening of $0.1$ for plotting purposes). At $J \!\lesssim\! 0.3$, $n_{z^2}$ has decreased sufficiently so that the $\Gamma$ pocket is lifted above $\omega \!=\! 0$. White lines show the bare Wannier bands.}
\label{fig:LaNiO2_Akw}
\end{figure*}

We now move on to a material-oriented setting and study the normal state of LaNiO$_2$.
While different models for LaNiO$_2$ with a variable number of orbitals have been constructed, 
we here consider a minimal two-orbital model \cite{Sakakibara2020,Hepting2020,Adhikary2020,Wan2021,Kang2021}
at fixed occupancy $N \!=\! 1$ (quarter-filling).
Evidently, this constitutes a strong simplification 
compared to more realistic simulations of infinite-layer nickelates
and can be seen as a
minimal extension beyond a model based on only the $d_{x^2-y^2}$ orbital.
Indeed, our model allows us to investigate to what extent orbital fluctuations 
and Hund physics in this compound persist when treated in a pure two-orbital description.

Henceforth, our unit of energy is $\mathrm{eV}$ and $J \!\equiv\! J_\mHK$.
We employ the two-orbital model of LaNiO$_2$ from Ref.~\cite{Kang2021} (see their Fig.~S16) constructed from the DFT band structure in a small energy window of roughly $[-1.5,2.5]$ with respect to the Fermi level.
Differently from other models \cite{Sakakibara2020,Hepting2020,Adhikary2020,Wan2021},
we thereby choose two Ni-centered Wannier functions \cite{Kang2021}.
Next to the strongly correlated orbital of mostly Ni $d_{x^2-y^2}$ character,
we have a weakly correlated hybridized orbital of both Ni and La $d_{3z^2-r^2}$ character ($d_{z^2}$ for short), acting as the self-doping band.
We label this orbital $\textrm{h-}d_{z^2}$ (hybridized $d_{z^2}$) to emphasize that it is not equivalent to an orbital of dominant Ni $d_{z^2}$ character (as obtained, e.g., from a large energy-window construction).
Our model has a sizable crystal-field splitting between the $d_{x^2-y^2}$ and $\textrm{h-}d_{z^2}$ orbital of 
$\Delta \!=\! \epsilon_{z^2} \!-\! \epsilon_{x^2-y^2} \!\approx\! 0.8$. 
We will see that, in the quarter-filled two-orbital system, this large $\Delta$ leads to a large orbital polarization between the $d_{x^2-y^2}$ and $\textrm{h-}d_{z^2}$ orbital (the former almost half-filled, the latter almost empty), suppressing orbital fluctuations and Hund-metal characteristics.

We choose $U \!=\! 3.1$ and $J \!=\! 0.7$, as in the narrow energy-window three-orbital setting of Ref.~\cite{Karp2020}.
Attempting a realistic description, one would have to modify these values:
the large spread of our 
$\textrm{h-}d_{z^2}$ orbital (see Fig.~S16 of Ref.~\cite{Kang2021})
entails a smaller $U$ value for this orbital
and a reduced $J$ value.
However, we here refrain from a very realistic description and aim for a minimal model with few parameters.
Test calculations with $U_{\textrm{h-}d_{z^2}} \!<\! U_{x^2-y^2}$ only enforced the difference in correlation strengths
($Z_{z^2} \!\gg\! Z_{x^2-y^2}$, see below), and reducing $J$ weakens the suppressed Hund physics (see below) even further.

Figure~\ref{fig:LaNiO2_spectrum}(a) displays the band structure of LaNiO$_2$ from DFT and from our two-orbital model.
The model reproduces the original band structure quantitatively within, say, $[-0.5,1]$, 
and qualitatively within $[-1,2]$.
Statements about LaNiO$_2$ at larger energies are evidently not captured by the model.
Figure~\ref{fig:LaNiO2_spectrum}(b) shows the local spectral functions $\Ac_{\mathrm{loc}}$ from the bare model (dashed  lines) and from DMFT (full lines).
As can be readily seen, correlations (incorporated via DMFT) change the spectrum significantly.
Especially in the $d_{x^2-y^2}$ orbital, the quasiparticle peak is sharpened and notable Hubbard side bands emerge.
We indicate the occupancy per orbital $n_m$ in brackets.
The orbital polarization ($n_{x^2-y^2} \!\gg\! n_{z^2}$) is slightly decreased by correlations but still sizable.
Since the $\textrm{h-}d_{z^2}$ orbital is barely occupied, it is much less correlated ($Z_{z^2} \!\gg\! Z_{x^2-y^2}$).
The precise quasiparticle weight of the $d_{x^2-y^2}$ orbital depends on its proximity to half-filling \cite{Si2020}.
With $n_{x^2-y^2} \!=\! 0.82$, we find $Z_{x^2-y^2} \!=\! 0.29$, consistently with Fig.~S1 of Ref.~\cite{Si2020} and somewhat similarly to Ref.~\cite{Karp2020} (there, $Z_{x^2-y^2} \!=\! 0.25$ at $n_{x^2-y^2} \!=\! 0.903$ in a three-orbital setting for NdNiO$_2$ with the same $U$ and $J$ values as used here). 

\begin{table}[b]
\centering
\begin{tabular}{c||c|c||c|c|c|c|c|c}
$J$ & $n_{x^2-y^2}$ & $Z_{x^2-y^2}$ & $p_{0,0}$ & $p_{1,1/2}$ & $p_{2,1}$ & $p_{2,0}$ & $p_{3,1/2}$ & $p_{4,0}$
\\ \hline
0.7  & 0.82 & 0.29 & 15.5\% & 69.4\% & 10.0\% & 4.7\% & 0.4\% & 0
\\ \hline
0.3  & 0.94 & 0.28 & 10.7\% & 78.7\% & 2.9\% & 7.6\% & 0.1\% & 0
\\ \hline
0  & 0.98 & 0.27 & 9.4\% & 81.3\% & 0.6\% & 8.7\% & 0 & 0 
\end{tabular}
\caption{LaNiO$_2$ results for varying $J$ at $U \!=\! 3.1$. Probabilities $p_{N,S^z}$ for impurity occupancies at different charge $N$ and spin $S$. Even at $J \!=\! 0.7$, the spin-triplet state has less weight (10.1\%) than in the five-orbital calculation of Ref.~\cite{Wang2020}.}
\label{tab:LaNiO2_J}
\end{table}

We now turn to the effect of Hund's coupling $J$.
At $J \!=\! 0$, the large crystal-field splitting would yield an almost depleted $\textrm{h-}d_{z^2}$ orbital,
whereas large $J$ promotes a large-spin state involving both orbitals.
Table~\ref{tab:LaNiO2_J} confirms this expectation, giving the values for $n_{x^2-y^2}$ 
(while $n_{z^2} \!=\! 1 \!-\! n_{x^2-y^2}$) for $J \!=\! 0.7, 0.3, 0$ at fixed $U \!=\! 3.1$.
As expected, $Z_{x^2-y^2}$ is reduced as the orbital approaches half-filling \cite{Si2020}.
In Fig.~\ref{fig:LaNiO2_Akw}, showing $\mathcal{A}({\vec{k},\omega})$,
one can see how the $\Gamma$ pocket is lifted above the Fermi level for sufficiently low $n_{z^2}$, realized at $J \!\lesssim\! 0.3$.
Table~\ref{tab:LaNiO2_J} also gives the eigenvalues of the impurity density matrix for each symmetry sector (i.e., the valence histogram \cite{Shim2007}).
Of particular interest is the weight $p_{2,1}$ of the high-spin multiorbital state. It strongly decreases with decreasing $J$.
However, even at large $J \!=\! 0.7$, $p_{2,1} \!=\! 10.1\%$ is much smaller than the value in Ref.~\cite{Wang2020}, treating LaNiO$_2$ with five correlated orbitals ($25.9\%$ in their Table~II; $Z_{x^2-y^2} \!=\! 0.36$ in their Table~I is comparable to ours).

\begin{figure}[t]
\includegraphics[scale=1]{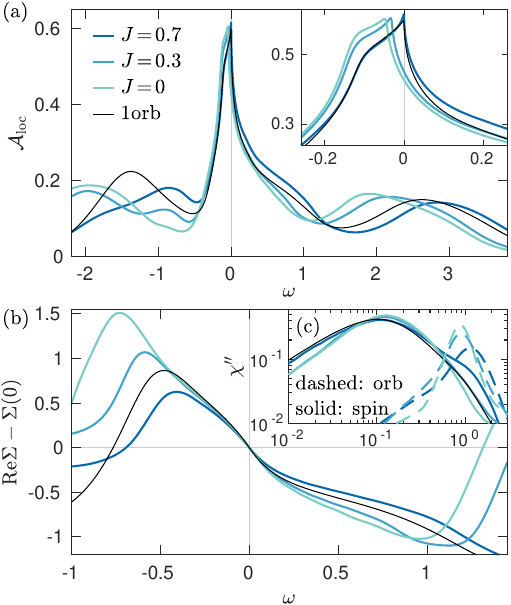}
\caption{%
(a) Spectral functions and (b) real part of self-energies of the $d_{x^2-y^2}$ orbital; (c) susceptibilities. The results stem from variable $J$ in the two-orbital model of LaNiO$_2$ and from a one-shot one-orbital (1orb) calculation starting from the two-orbital solution at $J \!=\! 0.7$.}
\label{fig:LaNiO2_dyn}
\end{figure}

Next, we scrutinize to what extent dynamic correlators of the more correlated $d_{x^2-y^2}$ orbital are affected by $J$.
We present in Fig.~\ref{fig:LaNiO2_dyn} the spectral functions, self-energies (real part), and spin and orbital susceptibilities for $J\!\in\!\{ 0.7, 0.3, 0\}$ at $U \!=\! 3.1$. 
We also include (as a black line) results from a one-shot one-orbital calculation, where we kept $U\!=\! 3.1$ and took the hybridization function for the $d_{x^2-y^2}$ orbital from the two-orbital $J\!=\! 0.7$ calculation.
For all values of $J$ and even the one-orbital calculation,
the low-energy behavior is very similar.
This includes the shape of the quasiparticle peak in $\Ac$, the slope in $\Re \Sigma$, and the peak position of $\chi''_\msp$ (as a measure for $T^\mK_\msp$).
Note that the peak of $\chi''_\morb$ occurs at higher energies than that of $\chi''_\msp$ even at $J \!=\! 0$ (trivial SOS due to large $\Delta$) and shifts only slightly to higher energies with increasing $J$.
This comparison demonstrates that the effect of $J$ in our two-orbital model for LaNiO$_2$ is mostly restricted to the orbital occupancies.
Indeed, as a one-orbital calculation with the appropriate input suffices to reproduce the relevant quantities at low energies \cite{Karp2020},
we confirm the interpretation that the low-energy physics of LaNiO$_2$ is that of a single-band Hubbard model
with an appropriate self-doping (provided by other orbitals) \cite{Karp2020,Kitatani2020}.

\begin{figure}[t]
\includegraphics[scale=1]{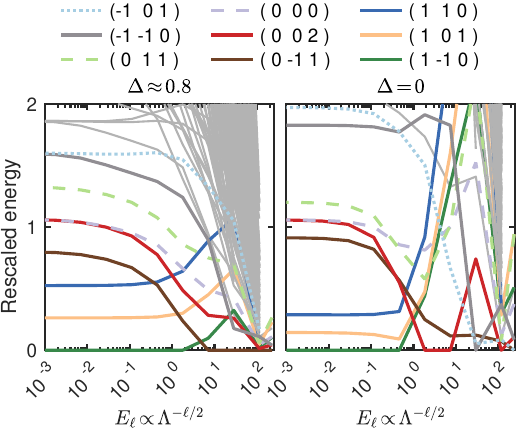}
\caption{%
NRG flow diagram for LaNiO$_2$ in the original setting $\Delta \!\neq\! 0$ and from a computational experiment where we manually set $\Delta \!=\! 0$.}
\label{fig:LaNiO2_flow}
\end{figure}

Finally, Fig.~\ref{fig:LaNiO2_flow} shows the NRG flow diagram of our two-orbital model of LaNiO$_2$ at $U \!=\! 3.1$, $J \!=\! 0.7$.
The left panel resembles the rightmost panel in Fig.~\ref{fig:Deltaflow} in that the size of the crystal-field splitting prevents 2SKS.
As a computational experiment, we add to our model an extra crystal field to achieve $\Delta \!=\! 0$,
thus compensating the original splitting.
The corresponding flow diagram (right panel in Fig.~\ref{fig:LaNiO2_flow}) now has an
intermediate region with a spin-triplet lowest-energy state that screens the orbital momentum.
This confirms that it is indeed the crystal-field splitting (or equivalently the polarization) that suppresses low-energy Hund physics in LaNiO$_2$.

\section{Conclusion}
\label{sec:conclusion}
We studied Hund physics in idealized two- and three-orbital systems and in a two-orbital model of LaNiO$_2$.
Using NRG as a DMFT impurity solver allowed us to examine high-resolution real-frequency spectra and the NRG flow from high to low energies. 
The latter was found to be a uniquely adequate tool to check for 2SKS,
the arguably most fascinating mechanism behind the SOS of Kondo scales---a hallmark of Hund metals.

First, we focused on orbital-symmetric models and compared three cases of $N$ electrons in $M$ orbitals:
$(M,N) \!\in\! \{ (3,2), (3,1), (2,1) \}$.
As indications of Hund physics, we examined
(i) the Janus effect in $Z(U,J)$,
(ii) the SOS of Kondo scales,
(iii) 2SKS in the RG flow with an intermediate orbital-singlet maximal-spin state,
and (iv) the orbital-resonance shoulder in $\mathcal{A}$ and the corresponding inverted slope in $\Re \Sigma$ (for orbital-symmetric models).
The case $(M,N) \!=\! (3,2)$ is known to prominently feature Hund physics:
Since $J$ splits the atomic ground-state manifold, $Z$ and $T^\mK_\msp$ decrease drastically with $J$,
enabling (i) a clear Janus effect and (ii) wide SOS.
(iii) The RG flow has a wide energy window between $T^\mK_\msp$ and $T^\mK_\morb$ dominated by an orbital-singlet maximal-spin state,
and (iv) $\mathcal{A}$ shows a pronounced shoulder ($\Re \Sigma$ an inverted slope).

Since $J$ does not affect the atomic ground state for $N=1$, properties (i) and (ii) are not very pronounced in the two cases $(M,N) \!\in\! \{(3,1), (2,1) \}$.
For $(M,N) \!=\! (3,1)$, one does not observe any signs of properties (iii) and (iv) either.
However, for $(M,N) \!=\! (2,1)$, even though the window between $T^\mK_\msp$ and $T^\mK_\morb$ is relatively small (slightly larger than a decade),
one can distinguish an orbital-singlet spin-triplet state as the lowest-energy state.
Similarly, one can distinguish a weak shoulder in $\mathcal{A}$ (inverted slope in $\Re \Sigma$).
Thereby, we substantiate the notion of a weak Hund metal
for $(M,N) \!=\! (2,1)$ \cite{Ryee2021}.
The RG flow reveals that 2SKS and the intermediate orbital-singlet maximal-spin state formed by binding a bath electron is facilitated by $N \!=\! M \!-\! 1$ (i.e., one-charge proximity to half-filling, independent of $N$), while it is unlikely for $N \!=\! M \!-\! 2$ (where two bath electrons would be required to form an orbital singlet).

We then studied the effect of a crystal-field splitting $\Delta$ and showed that 2SKS is obviated if $|\Delta| \!\gtrsim\! T^\mK_\morb(\Delta \!=\! 0)/2$.
In systems with strongly broken orbital symmetry, 2SKS does not go hand in hand with SOS or the shoulder in $\Ac$ (inverted slope in $\Re \Sigma$):
On the one hand, for large $\Delta$, $\chi''_\morb$ peaks at the bare parameter $\Delta$ due to atomic quenching of orbital fluctuations (and not dynamic Kondo screening).
On the other hand, the shoulder in $\Ac$ then changes in nature from an orbital-resonance shoulder to a separate doublon-holon peak \cite{Kugler2019}.

An important next step is to elucidate how 2SKS and the intermediate orbital-singlet maximal-spin state are reflected in observables; one candidate are fractional power laws at intermediate frequencies in local susceptibilities \cite{Walter2020,Wang2020a}.
Here, our observation of 2SKS in a quarter-filled two-orbital model is helpful as it constitutes a conceptually and numerically simpler setting than the previously known case $(M,N) \!=\! (3,2)$.
We thus suggest a similar study as in Refs.~\cite{Walter2020,Wang2020a}, using a generalized Kondo model to manually increase the intermediate energy window between $T^\mK_\msp$ and $T^\mK_\morb$, but for $(M,N) \!=\! (2,1)$.

Second, we analyzed a two-orbital model of LaNiO$_2$,
containing the strongly correlated $d_{x^2-y^2}$ orbital 
and a weakly correlated hybridized orbital of $d_{z^2}$ symmetry
(not equivalent to the Ni $d_{z^2}$ orbital of a large energy-window construction).
Our aim was to study a minimal setting in which to test Hund physics
while undoubtedly compromising on the degree of realism.
In particular, the frozen charge of one electron in the low-energy two-band window 
is a significant simplification---%
it is known that different orbital occupancies in different models lead to variable outcomes on quantities like $Z_{x^2-y^2}$.

In our quarter-filled two-orbital model, $J$ mainly affects the orbital occupancies 
while low-energy Hund physics are suppressed by the large $\Delta \!\approx\! 0.8\,\mathrm{eV}$.
This confirms the interpretation that the low-energy physics of LaNiO$_2$ is that of a single-band Hubbard model with an appropriate self-doping (provided by other orbitals) \cite{Karp2020,Kitatani2020}.
Nevertheless, we showed that multiorbital effects in our two-orbital model can be seen upon counteracting the crystal-field splitting, thus reducing the orbital polarization.
A first test within DFT showed that positive strain along the $c$ direction at constant volume of the unit cell (thus decreasing $c/a$, where $a \!=\! b$, and $c/a \!\approx\! 0.87$ in the pristine structure) decreases the energy splitting between the two orbitals.
It would be interesting to realize similar effects in experiment and thus potentially induce low-energy Hund physics.

\acknowledgments
We thank Sangkook Choi and Antoine Georges for fruitful discussions and Seung-Sup B.~Lee for a critical reading of the manuscript.
FBK and GK were supported by NSF Grant No.\ DMR-1733071,
CJK by NRF Grant No.~2022R1C1C1008200.
FBK acknowledges support by the Alexander von Humboldt Foundation through the Feodor Lynen Fellowship. 
The Flatiron Institute is a division of the Simons Foundation.

\appendix

\section{Comparing quasiparticle weights}
\label{app:more_results}
\begin{figure*}[t]
\includegraphics[scale=1]{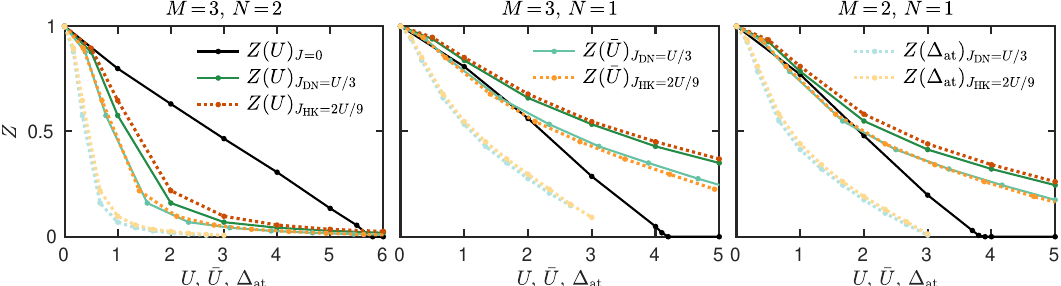}
\caption{%
Quasiparticle weight $Z$ as a function of $U$, $\bar{U}$, or $\Delta_{\mathrm{at}}$, analogous to Fig.~\ref{fig:Z_A_SE}(a), compared between the DN and the HK Hamiltonians. The behavior is qualitatively very similar for both Hamiltonians. Compared as a function of $\bar{U}$ or $\Delta_{\mathrm{at}}$, the results even agree on a quantitative level.}
\label{fig:Z}
\end{figure*}

Figure~\ref{fig:Z} shows that $Z$ depends very similarly on $U$ and $J$ for both the DN and the HK interaction Hamiltonians. For the closest connection between both situations, we match the intraorbital interaction $U$
as well as the atomic gap $\Delta_{\mathrm{at}} \!=\! U' \!-\! J$. 
Next to $U \!\equiv\! U_\mDN \!=\! U_\mHK$, this leads to the relation
\begin{align}
U'_\mDN \!-\! J_\mDN = U'_\mHK \!-\! J_\mHK
\Leftrightarrow
U \!-\! 2 J_\mDN = U \!-\! 3 J_\mHK
.
\end{align}
This simplifies to $J_\mHK \!=\! \frac{2}{3} J_\mDN$,
and $J_\mDN \!=\! \frac{1}{3}U \Rightarrow J_\mHK \!=\! \frac{2}{9}U$.

\section{Operators and interaction Hamiltonians}
\label{app:Hamiltonians}
In a multiorbital system, we have the (local) charge and spin operators
\begin{align}
N & = \sum_{m\sigma} d^\dag_{m\sigma} d_{m\sigma}
, \quad
\vec{S} = \tfrac{1}{2} \sum_{m\sigma\sigma'} d^\dag_{m\sigma} \vec{\sigma}_{\sigma\sigma'} d_{m\sigma'}
,
\end{align}
with the Pauli matrices $\vec{\sigma}$.
For $M \!=\! 3$, specifically, we have 
\begin{align}
T_a & = \tfrac{1}{2} \sum_{mm'\sigma} d^\dag_{m\sigma} \lambda^a_{mm'} d_{m'\sigma}
, \\
L_m & = \mi \sum_{m'm''\sigma} \epsilon_{mm'm''} d^\dag_{m'\sigma} d_{m''\sigma}
\end{align}
as orbital operators.
Here, $T_a$ ($a=1, \ndots, 8$) applies to SU(3) symmetry and uses the Gell-Mann matrices $\lambda^a$,
while $L_m$ ($m=1, \ndots, 3$) applies to SO(3) symmetry and uses the Levi-Civita symbol $\epsilon_{mm'm''}$.
For $M=2$, one can replace the Gell-Mann matrices by the Pauli matrices for SU(2) orbital symmetry,
and one can use $L \equiv L_3$ for SO(2) symmetry.

From these operators, we can build the quartic terms
\begin{align}
N^2, \quad \vec{S}^2, \quad \vec{T}^2, \quad \vec{L}^2
.
\label{eq:interactions_terms}
\end{align}
In the $S^z$ basis, one often refers to the following interactions
as \textbf{d}ensity-\textbf{d}ensity (intraorbital, interorbital opposite-spin, same-spin), \textbf{s}pin-\textbf{f}lip, and \textbf{p}air-\textbf{h}opping terms, respectively:
\begin{align}
H^{(1)}_{\mathrm{dd}} & = \!\sum_m n_{m\uparrow} n_{m\downarrow}
, \quad
H^{(2)}_{\mathrm{dd}} = \!\sum_{m\neq m'} n_{m\uparrow} n_{m'\downarrow}
, \nonumber \\
H^{(3)}_{\mathrm{dd}} & = \!\!\sum_{m<m',\sigma} \!\!\!\! n_{m\sigma} n_{m'\sigma}
, \quad
H_{\mathrm{sf}} = - \!\!\sum_{m\neq m'} \! d^\dag_{m\uparrow} d_{m\downarrow} d^\dag_{m'\downarrow} d_{m'\uparrow}
, \nonumber \\
H_{\mathrm{ph}} & = \!\sum_{m\neq m'} \! d^\dag_{m\uparrow} d^\dag_{m\downarrow} d_{m'\downarrow} d_{m'\uparrow}
.
\end{align} 
Their relation to the terms in Eq.~\eqref{eq:interactions_terms}, as derived below, is 
\begin{subequations}
\label{eq:interaction_terms_relations}
\begin{align}
& \tfrac{1}{2} N(N \!-\! 1) = H^{(1)}_{\mathrm{dd}} + H^{(2)}_{\mathrm{dd}} + H^{(3)}_{\mathrm{dd}} 
, \\
& \vec{S}^2 + \tfrac{1}{4}N^2 - N = -H^{(1)}_{\mathrm{dd}} + H^{(3)}_{\mathrm{dd}} - H_{\mathrm{sf}}
, 
\label{eq:tildeS2}
\\
& \vec{T}^2 + \tfrac{1}{2M} N^2 - \tfrac{M}{2} N
= H^{(1)}_{\mathrm{dd}} - H^{(3)}_{\mathrm{dd}} + H_{\mathrm{sf}}
, 
\label{eq:tildeT2}
\\
& \tfrac{1}{2} [ \vec{L}^2 - (M \!-\! 1) N ] = - H^{(3)}_{\mathrm{dd}} + H_{\mathrm{sf}} - H_{\mathrm{ph}}
\label{eq:tildeL2}
.
\end{align}
\end{subequations}
Evidently, $\vec{T}^2$ is linearly dependent on $N^2$, $N$, and $\vec{S}^2$.
We abbreviate Eqs.~\eqref{eq:tildeS2}--\eqref{eq:tildeL2} as $\tilde{\vec{S}}^2$, $\tilde{\vec{T}}^2$, and $\tfrac{1}{2}\tilde{\vec{L}}^2$, respectively.

The combinations forming the generalized Kanamori (GK) and the HK Hamiltonians are 
(cf.\ Ref.~\cite{Georges2013} for $M=3$)
\begin{align}
H_{\mathrm{GK}}
& = 
\frac{3U'-U}{2} \tfrac{1}{2}N(N \!-\! 1)
+
(U'-U) \vec{S}^2
\nonumber \\
& \
+
(U'-U+J) \tfrac{1}{2} \vec{L}^2
+
\big[ \tfrac{1+2M}{4}(U-U') + \tfrac{1-M}{2} J \big] N
\nonumber \\
& =
U' \tfrac{1}{2}N(N \!-\! 1)
+
(U'-U) \tilde{\vec{S}}^2
+
(U'-U+J) \tfrac{1}{2} \tilde{\vec{L}}^2
\nonumber \\
& =
U H^{(1)}_{\mathrm{dd}} + U' H^{(2)}_{\mathrm{dd}} + (U'-J) H^{(3)}_{\mathrm{dd}} + J H_{\mathrm{sf}}
\nonumber \\
& \ 
+ (U-U'-J) H_{\mathrm{ph}}
,
\end{align}
which, for $U' = U - 2J$, simplifies to
\begin{align}
H_{\mathrm{HK}}
& =
(U-3J) \tfrac{1}{2}N(N \!-\! 1)
- 2J \vec{S}^2
- J \tfrac{1}{2} \vec{L}^2
+ \tfrac{2+M}{2} J N
\nonumber \\
& =
U' \tfrac{1}{2}N(N-1)
-2J \tilde{\vec{S}}^2
-J \tfrac{1}{2} \tilde{\vec{L}}^2
\nonumber \\
& =
U H^{(1)}_{\mathrm{dd}} + U' H^{(2)}_{\mathrm{dd}} + (U' \!-\! J) H^{(3)}_{\mathrm{dd}}
+ J (H_{\mathrm{sf}} \!+\! H_{\mathrm{ph}})
.
\end{align}
The DN Hamiltonian is given by \cite{Georges2013}
\begin{align}
H_{\mathrm{DN}}
& =
(U - \tfrac{3}{2}J) \tfrac{1}{2}N(N \!-\! 1)
- J \vec{S}^2
+ \tfrac{3}{4} J N
\nonumber \\
& = (U - J) \tfrac{1}{2}N(N \!-\! 1)
- J \tilde{\vec{S}}^2
\nonumber \\
& = (U - J) \tfrac{1}{2}N(N \!-\! 1)
- 2 J \tilde{\vec{S}}^2
- J \tilde{\vec{T}}^2
\nonumber \\
& =
U H^{(1)}_{\mathrm{dd}} + (U \!-\! J) H^{(2)}_{\mathrm{dd}} + (U \!-\! 2J) H^{(3)}_{\mathrm{dd}}
+ J H_{\mathrm{sf}}
.
\end{align}
The second-to-last row is a trivial rewriting,
emphasizing the similarity to the HK Hamiltonian with prefactors $-2J$ and $-J$ in front of the spin and orbital operators squared, respectively.

To derive Eqs.~\eqref{eq:interaction_terms_relations}, we first normal-order the interactions:
\begin{align}
N^2
& = 
\sum_{\alpha\alpha'} d^\dag_{\alpha} d_{\alpha} d^\dag_{\alpha'} d_{\alpha'}
= 
\sum_{\alpha\alpha'} d^\dag_{\alpha} d^\dag_{\alpha'} d_{\alpha'} d_{\alpha} 
+ N
,
\label{eq:simplify_N2}
\end{align} 
where $\alpha \!=\! (m, \sigma)$.
Further, we have
\begin{align}
\vec{S}^2
& =
\tfrac{1}{4} \sum_{\substack{m\sigma\sigma' \\ \tilde{m}\tilde{\sigma}\tilde{\sigma}'}}
d^\dag_{m\sigma} d_{m\sigma'} d^\dag_{\tilde{m}\tilde{\sigma}} d_{\tilde{m}\tilde{\sigma}'} 
\vec{\sigma}_{\sigma\sigma'} \cdot \vec{\sigma}_{\tilde{\sigma}\tilde{\sigma}'}
.
\end{align}
Using 
$\vec{\sigma}_{\sigma\sigma'} \cdot \vec{\sigma}_{\tilde{\sigma}\tilde{\sigma}'} =
2\delta_{\sigma\tilde{\sigma}'}\delta_{\sigma'\tilde{\sigma}} - \delta_{\sigma\sigma'}\delta_{\tilde{\sigma}\tilde{\sigma}'}$,
we get
\begin{align}
\vec{S}^2
& =
\tfrac{1}{2} \sum_{mm'\sigma}
d^\dag_{m\sigma} d_{m\sigma'} d^\dag_{m'\sigma'} d_{m'\sigma} 
-
\tfrac{1}{4} N^2
.
\end{align}
Then,
$d_{m\sigma'} d^\dag_{m'\sigma'}
=
-d^\dag_{m'\sigma'} d_{m\sigma'}
+
\delta_{mm'}$ 
yields
\begin{align}
\vec{S}^2
=
\tfrac{1}{2}
\sum_{mm'\sigma\sigma'}
d^\dag_{m\sigma} d^\dag_{m'\sigma'} d_{m'\sigma} d_{m\sigma'} 
+ N - 
\tfrac{1}{4} N^2
.
\label{eq:simplify_S2}
\end{align}

In the case of $\mathrm{SU}(2)$ orbital symmetry, one finds
\begin{flalign}
\vec{T}^2
& \overset{M=2}{=}
-
\tfrac{1}{2}
\sum_{mm'\sigma\sigma'}
d^\dag_{m\sigma} d^\dag_{m'\sigma'} d_{m'\sigma} d_{m\sigma'} 
+ N - 
\tfrac{1}{4} N^2
\hspace{-0.5cm}
&
\label{eq:simplify_T2_M=2}
\end{flalign}
by a simple replacement of $m \!\leftrightarrow\! \sigma$.
For $M \!=\! 3$, we have
\begin{align}
\vec{T}^2
& =
\tfrac{1}{4} \sum_{\substack{mm'\sigma \\ \tilde{m}\tilde{m}'\tilde{\sigma}}}
d^\dag_{m\sigma} d_{m'\sigma} d^\dag_{\tilde{m}\tilde{\sigma}} d_{\tilde{m}'\tilde{\sigma}} 
\sum_a \lambda_{mm'} \lambda_{\tilde{m}\tilde{m}'}
.
\end{align}
Using
$\sum_a \lambda_{mm'} \lambda_{\tilde{m}\tilde{m}'}
=
2\delta_{m\tilde{m}'}\delta_{m'\tilde{m}} - \tfrac{2}{3}\delta_{mm'}\delta_{\tilde{m}\tilde{m}'}$,
we get
\begin{align}
\vec{T}^2
& =
\tfrac{1}{2} \sum_{mm'\sigma\sigma'}
d^\dag_{m\sigma} d_{m'\sigma} d^\dag_{m'\sigma'} d_{m\sigma'} 
-
\tfrac{1}{6} N^2
.
\end{align}
We commute
$d_{m'\sigma} d^\dag_{m'\sigma'}
=
-d^\dag_{m'\sigma'} d_{m'\sigma}
+ \delta_{\sigma\sigma'}$
to obtain
\begin{flalign}
\vec{T}^2
& =
-
\tfrac{1}{2} \sum_{mm'\sigma\sigma'}
d^\dag_{m\sigma} d^\dag_{m'\sigma'} d_{m'\sigma} d_{m\sigma'} 
+
\tfrac{3}{2} N 
-
\tfrac{1}{6} N^2
.
\hspace{-0.5cm}
&
\label{eq:simplify_T2_M=3}
\end{flalign}
We can thus summarize both cases $M \!\in\! \{ 2, 3 \}$ as
\begin{align}
\vec{T}^2
& =
-
\tfrac{1}{2} \sum_{mm'\sigma\sigma'}
d^\dag_{m\sigma} d^\dag_{m'\sigma'} d_{m'\sigma} d_{m\sigma'} 
+ \tfrac{M}{2} N - \tfrac{1}{2M} N^2 
.
\label{eq:simplify_T2}
\end{align}

For $\mathrm{SO}(M)$ orbital symmetry, we have
\begin{align}
\vec{L}^2
& =
- \sum_{\substack{kl\sigma \\ k'l'\sigma'}}
\sum_m \epsilon_{mkl} \epsilon_{mk'l'}
d^\dag_{k\sigma} d_{l\sigma} d^\dag_{k'\sigma'} d_{l'\sigma'}
,
\end{align}
where
$m,k,l,k',l' \in \{1,2,3\}$ for $M=3$ and
$k,l,k',l' \in \{1,2\}$ while $m=3$ for $M=2$.
In both cases, we can use
$\sum_m \epsilon_{mkl} \epsilon_{mk'l'}
=
\delta_{kk'}\delta_{ll'} - \delta_{kl'}\delta_{lk'}$
to get
\begin{align}
\vec{L}^2
& =
\sum_{kl\sigma\sigma'}
(
d^\dag_{k\sigma} d_{l\sigma} d^\dag_{l\sigma'} d_{k\sigma'}
-
d^\dag_{k\sigma} d_{l\sigma} d^\dag_{k\sigma'} d_{l\sigma'}
)
.
\end{align}
Via
$d_{l\sigma} d^\dag_{l\sigma'}
=
-d^\dag_{l\sigma'} d_{l\sigma}
+ \delta_{\sigma\sigma'}$
and
$d_{l\sigma} d^\dag_{k\sigma'}
=
-d^\dag_{k\sigma'} d_{l\sigma}
+ \delta_{kl}\delta_{\sigma\sigma'}
$,
we find
\begin{align}
\vec{L}^2
& =
\sum_{mm'\sigma\sigma'}
(
d^\dag_{m\sigma} d^\dag_{m'\sigma'} d_{m\sigma'} d_{m'\sigma} 
-
d^\dag_{m\sigma} d^\dag_{m\sigma'} d_{m'\sigma'} d_{m'\sigma} 
)
\nonumber \\
& \
+ (M-1)N
.
\label{eq:simplify_L2}
\end{align}

Now, Eq.~\eqref{eq:simplify_N2} is simplified,
after dividing $\sum_{mm'\sigma\sigma'}$ into
$\sum_{m=m',\sigma\neq\sigma'}$,
$\sum_{m\neq m',\sigma=\sigma'}$, and
$\sum_{m\neq m',\sigma\neq\sigma'}$, as
\begin{align}
\tfrac{1}{2} N(N \!-\! 1)
& =
\tfrac{1}{2}
\sum_{mm'\sigma\sigma'}
d^\dag_{m\sigma} d^\dag_{m'\sigma'} d_{m'\sigma'} d_{m\sigma}
\nonumber \\
& =
H^{(1)}_{\mathrm{dd}} + H^{(2)}_{\mathrm{dd}} + H^{(3)}_{\mathrm{dd}}
.
\end{align}
Similarly, Eqs.~\eqref{eq:simplify_S2}, \eqref{eq:simplify_T2}, and \eqref{eq:simplify_L2} give
\begin{align}
\tilde{\vec{S}}^2 
& =
- \tilde{\vec{T}}^2 
=
\tfrac{1}{2}
\!\!\sum_{mm'\sigma\sigma'}\!\!
d^\dag_{m\sigma} d^\dag_{m'\sigma'} d_{m'\sigma} d_{m\sigma'}
\nonumber \\
& =
-H^{(1)}_{\mathrm{dd}} + H^{(3)}_{\mathrm{dd}} - H_{\mathrm{sf}}
,
\\
\tfrac{1}{2} \tilde{\vec{L}}^2
& =
\tfrac{1}{2} \sum_{mm'\sigma\sigma'}
d^\dag_{m\sigma} 
( d^\dag_{m'\sigma'} d_{m\sigma'}
-
d^\dag_{m\sigma'} d_{m'\sigma'} )
d_{m'\sigma} 
\nonumber \\
& =
- H^{(3)}_{\mathrm{dd}} + H_{\mathrm{sf}} - H_{\mathrm{ph}}
.
\end{align}
Thus, Eqs.~\eqref{eq:interaction_terms_relations} are derived.
For completeness, we mention that
$\vec{S}^2$ can be decomposed into
$\sum_m \vec{S}_m^2 \!=\! -\tfrac{3}{2} H^{(1)}_{\mathrm{dd}} + \tfrac{3}{4} N$ and
$2\sum_{m<m'} \vec{S}_m \cdot \vec{S}_{m'} \!=\!
- \tfrac{1}{2} H^{(2)}_{\mathrm{dd}} + \tfrac{1}{2} H^{(3)}_{\mathrm{dd}} - H_{\mathrm{sf}}$.

\bibliography{references}

\end{document}